\documentclass[trackchanges]{aastex63}
\usepackage{mathtools,upgreek}

\received{6 July 2020}
\revised{30 July 2020}
\accepted{3 August 2020}
\submitjournal{Planetary Science Journal}

\shorttitle{}

\newcommand{\pennstate}{Department of Astronomy \& Astrophysics and \\
Center for Exoplanets and Habitable Worlds and \\ 
Penn State Extraterrestrial Intelligence Center \\
525 Davey Laboratory \\ The Pennsylvania State University \\ University Park, PA, 16802, USA}

\begin{document}

\title{Barycentric Corrections for Precise Radial Velocity Measurements of Sunlight}

\correspondingauthor{Jason Wright}
\email{astrowright@gmail.com}

\author[0000-0001-6160-5888]{Jason T.\ Wright}
\affil{\pennstate}

\author[0000-0001-8401-4300]{Shubham Kanodia}
\affil{\pennstate}

\begin{abstract}

We provide formulae for the calculation of precise Doppler velocities of sunlight, in both the case of direct observations of the Sun and in reflection from the surfaces of solar system objects such as the Moon or asteroids. We discuss the meaning of a ``barycentric correction'' of measurements of these Doppler velocities, which is a different procedure from the analogous correction for starlight, and provide a formula for reducing such measurements to the component of the Sun's motion in the direction of the Earth or other Solar System object. We have implemented this procedure in the public {\tt barycorrpy} \texttt{Python} package, and use it to explore the properties of the barycentric-corrected Doppler velocity of sunlight over 30 years. When measured directly, we show it is dominated by the non-periodic motion due to Jupiter, and that the signals of the other planets, including Venus, are not discernible in Fourier space. We show that ``detecting'' Venus in the Doppler velocities of sunlight will require either observing sunlight in reflection from an asteroid, or modeling the their individual contributions to the motion of the Sun in counterfactual kinematic or dynamical simulations of the Solar System with and without them.

\end{abstract}


\section{Introduction} \label{sec:intro}

\subsection{The Value of Solar RV Measurements}

In extremely precise radial velocity (EPRV) work, the Sun is our only touchstone. We know its center of mass motion with respect to our observatories essentially perfectly, and we can resolve the surface spatially and spectrally with exquisite precision. 

We can also determine the amount of variation the Sun {\it should} exhibit at any given moment by integrating disk-integrated Dopplergrams of a single spectral line to synthesize the sorts of observations we would make with our spectrographs, thus connecting the variability we observe in our spectrographs to physical phenomena we understand on the Sun. We can also use numerical models of the Sun validated against observations of it to interpret EPRV measurements of it and other stars \citep[e.g.][]{Cegla13,Cegla18}. The Sun is thus the ultimate RV ``standard star.'' 

These observations of the Sun-as-a-star have been used to great profit, teaching us much about the limitation of EPRV work on Sun-like-stars. For instance, \citet{Milbourne19} compared precise radial velocities with  data from the Solar Dynamics Observatory and SORCE satellite to determine the degree to which surface magnetic activity was causing RV variations. 

But observing the Sun as a star is generally challenging: it is too large and too bright to collect its disk-integrated light into EPRV spectrographs without dedicated hardware. Several spectrographs, including HARPS, HARPS-N, and NEID spectrographs make good use of small solar telescope feeds to make observations of the Sun during the day \citep[e.g][]{Collier19}.

Observations of the Sun at night \citep{McMillan93,Molaro11,Haywood16} are an elegant solution to the problem: the surfaces of the Moon and asteroids are good diffusive reflectors, providing a bright source of disk-integrated sunlight that can be observed with the very telescopes and optical fibers we use to observe stars. When observed this way, these sources are also immune to the effects of differential extinction across the solar disk that infect observations with solar telescopes, and their albedos change sufficiently slowly with wavelength that their reflection spectra will not interfere with EPRV measurements. 

Indeed, \citet{Haywood16} used simultaneous observations of Vesta and the Sun via the Solar Dynamics Observatory to deduce the physical origins of EPRV ``jitter'' on the sun and test the value of chromospheric activity indicators in correcting for it. \citet{Lanza16} used radial velocity measurements of the Sun in reflection to characterize the effects of the 11-year solar cycle on long-term EPRV measurements.

 \subsection{Challenges of Applying a ``Barycentric Correction'' to the Measured Doppler Shift of Sunlight}

\label{sec:challenges}
Precise Doppler work requires careful correction for the motion of the observatory in the form of a ``barycentric correction.''  In coarse work with stars, this means subtracting the velocity of the telescope (due to the orbital and rotational motion of the Earth) in the direction of the star from the measured velocity to get the star's velocity with respect to the solar system barycenter.

Thanks to precise Solar System ephemerides from the Jet Propulsion Laboratory \citep{DE430}, tracking of the Earth's orientation in space by the International Earth Rotation Service, and precise observatory coordinates from the Global Positioning System, the velocity vector of any observatory through the Solar System can be computed quickly and easily.  Applying this information to transform measured redshifts into radial velocity measurements is not trivial, however. The various aspects of the problem---including interpretation of the JPL ephemeris, handling perspective effects towards target stars, defining the effective time of observation in the presence of atmospheric effects, accounting for relativistic effects, and quantifying the uncertainties in the calculations---have been solved by multiple authors to the precision required for EPRV work on stars \citep[e.g.][]{Lindegren03,bary,BARYCORRPY,WrightRNAASbary,Feng19,Blackman19,Tronsgaard19}.

However, one cannot simply enter the Sun's sky coordinates into typical barycentric correction routine of the sort used by stellar astronomers and receive an easily interpreted result for a few reasons. Indeed, the concept of a ``barycentric correction'' for the Sun is not as clean or straightforward as the analogous concept for a star. 

One complication is that the radial direction to the Sun is highly variable, and so many of the approximations used in such a barycentric correction routine are invalid.  For instance, with other stars, the radial direction is approximately constant and changes in perspective cause small changes to the radial direction that are be treated as linear perturbations. For the Sun, the parallax is effectively $\pi$ radians, not a fraction of an arcsecond, and so the ``radial'' direction is not even approximately constant.  If one were to pick a constant ``radial'' vector for the Sun (say, the Sun's direction at the time of the first observation in a time series) then 3 months later one's measurements would contain almost no information about that component of the Sun's motion at all!

Another complication is that the purpose of making a barycentric correction is different for the Sun than with stars, so a proper philosophy of the calculation is unclear \citep[see Section 3.2 of][]{bary}. Here we are concerned with measuring the Sun's motion in the context of planet detection, in which case it is useful to imagine detecting, say, Jupiter via its gravitational influence on the Sun. But upon more careful consideration it is not obvious what this means precisely.

In a stellar context, it means comparing two models: one in which a star has the planet in question and one in which it does not.  For the first planet discovered orbiting a single star, this means comparing the null hypothesis of a star moving with constant velocity through space to one in which it is orbited by a single planet.  In this case, the barycentric correction procedure is clear, since for the small perturbations caused by a planet, the problems of calculating the Earth's motion and the star's motion due to a planet are separable.

In the case of the Solar System, we know the Sun is orbited by planets, so strictly speaking we are dealing with a counterfactual, not a null hypothesis. But what is that counterfactual, exactly? Jupiter and the other planets accelerate the Earth and our telescopes as much as or more than they accelerate the Sun.  Is our counterfactual that the other planets do not exist, so the Earth is a test particle on an unperturbed Keplerian orbit? Is that that the Earth's motion is influenced by them, but the Sun is not? When we write that the Sun is not influenced by them, do we only mean that we pretend that the Sun's Doppler shift is zero in the frame of the Solar System barycenter, or also that it sits at the barycenter of the Solar System? If the latter, how do we define the ``radial'' direction---from the observatory towards the apparent position of the solar disk center, or towards the Solar System barycenter?  And so on.

We will see, however, that we can define a ``barycentric correction'' in the sense of the correction we must apply to a measurement to get the redshift we would measure in the barycentric frame in the direction of the object that sunlight illuminates. This is not a ``radial velocity'' as measured along a constant direction, but from the constantly (and non-uniformly) varying direction from the Earth or other Solar System object to the apparent Sun.  The effects of the planets will thus not be that of true Keplerian orbital curves, and their effects on the Sun will not be concentrated at a single frequency and its harmonics.

It is perhaps conceptually simpler and more useful, then, to approach the problem with a better-defined question: what is the Doppler shift one should measure from a given observatory of a source of known frequency at the surface of the Sun?  This can be computed using much of the machinery of a barycentric correction---a Solar System ephemeris, Earth rotation parameters, accounting for relativistic effects---along with a few extra features unique to the problem (accounting for the finite speed of light and computing the retarded values of the position and velocity of the Solar System objects used in reflection). Any variations from this expected value can then be attributed to other complexities of the problem such as the finite size of the Sun, stellar jitter, or instrumental effects.

While this problem has been solved by multiple groups, there has not previously been a publication that describes a standard method or analyzes the precision possible for such work with existing ephemerides. In this article we explicate a rigorous barycentric correciton algorithm for precise Doppler observations of sunlight and describe public code that performs it. In Section~\ref{sec:theory} work we describe the theory of calculating Doppler shifts from ephemerides, and in Section~\ref{sec:reflection} we apply that theory to sunlight measured direction and in reflection.  Section~\ref{code} describes our public code implementation, and Section~\ref{sec:precision} describes how uncertainties propagate from the JPL ephemerides to the barycentric correction. In Section~\ref{sec:detecting} we use our code to simulate ``perfect'' observations of the Sun to assess the difficulty of ``detecting'' the Solar System planets in such data.

The term ``barycentric correction'' is sometimes also used in the context of defining {\it when} an observation has taken place, to refer to the reduction of the local time of observation to the time when the starlight of the observation passed the Solar System Barycenter. Observations of the Sun should instead be reduced to the time the light was {\it emitted} from the Sun, defined below as $t_e$ (see Section~\ref{sec:retarded}).

In this article we follow the notation of \citet{bary}. All positions, velocities, and times are measured in the frame of the Solar System Barycenter (SSB). We make use of the JPL Horizons interface \citep{JPLHorizons} which makes use of the DE series of ephemerides \citep{DE430}.

\section{Direct Observations of the Sun}

\label{sec:theory}

\subsection{Rigorous Formula for a Point Source}

The simplest case is observing the Sun directly.  We begin simply, considering a single point on the surface of the Sun, so we can use Equations 1 and 3 of \citet{bary}:
\begin{equation}
    z \equiv \frac{\nu_{\rm emit} - \nu_{\rm meas}}{\nu_{\rm meas}}
\end{equation}
\begin{equation}
    \nu_{\rm meas} = \nu_{\rm emit}\left(\frac{(1+z_{{\rm GR},\odot})}{\gamma_\odot (1+\vec{\beta}_\odot\cdot \hat{\rho}_{\odot,\earth})}\right)\left(\frac{\gamma_\earth (1+\vec{\beta}_\earth\cdot \hat{\rho}_{\odot,\earth})}{(1+z_{{\rm GR},\earth})}\right)\label{numeas}
\end{equation}

\noindent where the $\earth$ symbol stands for the observatory and $\odot$ for the Sun, and $z_{\rm GR}$ is the gravitational redshift from infinity to the observatory or Solar surface.\footnote{i.e., $z_{\rm GR}$ is a negative quantity. We maintain this convention from Equation 2 of \citet{bary}, which describes the {\it blueshift} of stellar photons falling into the solar gravitational well from infinity to the surface of the sun or to the observatory.} Heuristically, the first term in parentheses in Equation~\ref{numeas} shifts the frequency of light from the Sun's rest frame into that of the SSB (at infinity), and the second shifts it from the frame of the SSB into that of the observatory.\footnote{The classical Doppler shift calculation employs Galilean addition of velocities and Newtonian spacetime, and so can be expressed in terms of the relative velocity $(\vec{\beta}_\odot-\vec{\beta}_\earth)$; the relativistic expression in any given frame does not have this property. If one were to transform Eq.~\ref{numeas} to the frame of the source or observer, one would not only need to perform a relativistic addition of velocities, but also adjust the $\hat{\rho}_{\odot,\earth}$ unit vector to correct for aberration. The final answer is, of course, the same regardless of the inertial frames chosen, but the calculation in the frame of the SSB is both conceptually simplest and best connected to data in a Solar System ephemeris.}  Because of the light travel time in the system, we need to define {\it when} we are measuring these quantities.  Specifically, let $t$ be the time of observation, and $t_e$ be the time the photons were emitted. Then we have
\begin{eqnarray}
    \vec{\beta}_\earth &\equiv& \vec{v}_\earth(t)/c \\
    \vec{\beta}_\odot &\equiv& \vec{v}_\odot(t_e)/c \\
    \vec{\rho}_{\odot,\earth} &\equiv& \vec{r}_\odot(t_e) - \vec{r}_\earth(t) \\
    \rho &\equiv& |\vec{\rho}| \\
    \hat{\rho} &\equiv& \vec{\rho}/ \rho \\
    z_{GR,\odot} &=& - \frac{GM_{\odot}}{c^2 R_{\odot}} \label{zgrsun}\\
    z_{GR,\oplus} &=& -\frac{GM_{\odot}}{c^2 \rho_{\odot,\earth}} - \frac{GM_{\oplus}}{c^2 R_{\oplus}} \label{zgrearth}
\end{eqnarray}
\noindent where $\vec{r}$ and $\vec{v}$ refer to the position and velocity of an object, respectively, with respect to the SSB at some time.\footnote{Formally, equations \ref{zgrsun} and \ref{zgrearth} should include contributions from all Solar System bodies, including Jupiter and the Moon, however these only contribute at a level below 1 mm s$^{-1}$, and are time variable at a level far below that, so can be neglected in EPRV work.}  $\gamma$ is the usual Lorentz factor computed from the corresponding $\beta$:
\begin{equation}
    \gamma \equiv \frac{1}{\sqrt{1-\beta^2}}
\end{equation}

We measure the redshift of the spectral features of sunlight with respect to some fiducial set of frequencies $\nu^\prime$ that depend on the details of the measurement technique.  We thus measure

\begin{eqnarray}
    z_{\rm meas} &=& \frac{\nu^\prime - \nu_{\rm meas}}{\nu_{\rm meas}}\\
    &=& \frac{\nu^\prime}{\nu_{\rm emit}}\frac{(1+z_{{\rm GR},\earth})\gamma_\odot (1+\vec{\beta}_\odot\cdot \hat{\rho}_{\odot,\earth})}{(1+z_{{\rm GR},\odot})\gamma_\earth (1+\vec{\beta}_\earth\cdot \hat{\rho}_{\odot,\earth})}-1 \label{zmeas_sun}
\end{eqnarray}
\noindent from our source.

Here we encounter a few other differences with the usual barycentric correction. We see that the scale factor $\nu^\prime/\nu_{\rm meas}$ matters because we are interested in the direct connection between measured redshift and the precise, absolute motion of the Sun. (In exoplanet work this factor is degenerate with the imprecisely-known gravitational redshift, convective blueshift, and space motion of the star). 

Now we can, following the notation of \cite{bary}, define
\begin{equation}
    z_{\rm true} = \frac{\nu^\prime}{\nu_{\rm emit}}\frac{\gamma_\odot (1+\vec{\beta}_\odot\cdot \hat{\rho}_{\odot,\earth})}{(1+z_{{\rm GR},\odot})} - 1
\end{equation}
\noindent which captures all of the information about the redshift of the light of the Sun in its apparent direction as it would be observed from infinity, including the mean convective blueshift (captured in the $\nu^\prime/\nu_{\rm emit}$ term) and the gravitational redshift. This last term produces a $\sim634$ m s$^{-1}$ offset from zero for all measurements compared to typical values observed at Earth.\footnote{Including a $(1+z_{{\rm GR},\earth})$ term to put barycentric-corrected values near zero would introduce an annual signal from the Earth's motion in and out of the Sun's gravitational well due to its orbital eccentricity.} 

Then we can define a ``barycentric correction'' $z_B$ that transforms our measured redshifts to this value
\begin{equation}
    z_B = \frac{\gamma_\earth (1+\vec{\beta}_\earth\cdot \hat{\rho}_{\odot,\earth})}{(1+z_{{\rm GR},\earth})}-1 \label{zb}
\end{equation}
\noindent such that, following Equation~10 of \citet{bary} \begin{equation} 
    (1+z_{\rm true}) = (1+z_{\rm meas})(1+z_B) \label{corrected}
\end{equation}

One thus has two paths to ``correcting'' one's measurements of the Doppler shift of the sun ($z_{\rm meas}$) for barycentric motions.  The first is to model one's data using Eq.~\ref{zmeas_sun}, in which case the residuals will be due to measurement uncertainties, instrumental effects, and solar atmospheric effects (i.e., stellar jitter). 

If, instead, one wishes to remove the effects of the Earth's motion but retain the solar motion due to the planets, one can calculate $z_B$ using Eq.~\ref{zb}, and transform the measured redshifts to $z_{\rm true}$ using Eq.~\ref{corrected}. This should reveal, for instance, the $\sim10$ m/s signal of Jupiter at its synodic period with Earth of 1.09 years.

The $\nu^\prime/\nu_{\rm emit}$ term cannot be computed from an ephemeris and depends on the details of the measurement technique and the motions in the solar atmosphere.  In EPRV work, the templates or masks used to measure Doppler motions are usually chosen to keep this quantity as close to 1 as possible by setting the wavelengths of spectral features to those observed in sunlight.  This means that for sunlight, the ratio will differ from 1 by no more than the radial velocity of an observatory towards the sun in terms of the speed of light, which is of order $10^{-6}$. For computational purposes we therefore set it to 1, which will result in a small, constant, multiplicative offset between the predicted and measured differential in redshifts. It will also produce errors in the barycentric-corrected velocities via cross terms with $\beta_\odot$, which has amplitude of $4\times10^{-8}$ due to Jupiter. These cross terms therefore have magnitude $4\times10^{-14}$, corresponding to barycentric correction errors of order 20$\upmu s^{-1}$.

\subsection{(Un)importance of light travel time and finite source effects}\label{sec:solarLTT}

It is not strictly necessary to consider the retarded position of the Sun here. The error in the computed Doppler shift from neglecting the finite speed of light scales\footnote{A more rigorous upper bound on the error is $v_1+v_2$ where $v_1$ and $v_2$ are the speeds of the observer and source, times the error in the angle to the source, $ v_2\Delta t/d$ where $d$ is the distance to the source. Since  $\Delta t=d/c$ is the light travel time, the upper bound is thus $(v_1+v_2)\beta_2$} as $v_\earth v_\odot /c$. Since the Sun moves at only $\sim 10$ m/s, only at work below 1 cm/s would such effects begin to matter (retarded times will be important for other Solar System objects, though).

The real Sun has finite size, and so is not observed at (the retarded position of) its center of mass but at a range of angles and distances across its near hemisphere.  This has important consequences when considering the effects of, say, differential atmospheric extinction across the disk, which will preferentially favor photons from one limb of the Sun, with its characteristic rotational velocity and barycentric correction.

The largest (first order) effect of the varying barycentric correction comes from the tangential orbital motion of the Earth, whose projection in the direction of the Sun is in the form of a gradient across the disk with amplitude 150 m/s, which is one order of magnitude smaller than the gradient due to solar rotation. Like solar rotation, the average of this correction across the disk, neglecting surface features and atmospheric extinction, is zero, so this effect need not be part of our calculation. The second order effect is the decrease in the radial component of the velocity of the observatory towards the Sun as one observes away from disk center. This term does not average out but is smaller by a factor of $10^4$, so matters perhaps only at or below the 1 cm/s level. 

\section{Observations of the Sun in Reflection}
\label{sec:reflection}

\subsection{Formulae for Solar System Objects}

Solar System objects are essentially moving mirrors.  We must calculate the Doppler shift they impose on light at the time the light strikes them. This requires knowing the location and velocity of the light at the time it was emitted, the location and velocity of the surface it strikes at the time it strikes it, and the location and velocity of the primary mirror of the telescope at the time the light arrives there. 

Let $t_m$ be the time that the sunlight strikes the surface of a Solar System object, located at position $\vec{r}_m$.  The object moving at speed $\vec{\beta}_m$ sees the Sun in direction $\hat{\rho}_{\odot,m}$ as it was at time $t_e$, and at time $t$ the observatory sees the Solar System object in direction $\hat{\rho}_{m,\earth}$ as it was at time $t_m$, where
\begin{eqnarray}
    \vec{\beta}_m &\equiv& \vec{v}_m(t_m)/c \\
    \vec{\rho}_{\odot,m} &\equiv& \vec{r}_\odot(t_e) - \vec{r}_m(t_m)\\
    \vec{\rho}_{m,\earth} &\equiv& \vec{r}_m(t_m) - \vec{r}_\earth(t) 
\end{eqnarray}

For an observer on the Solar System object, we can calculate the observed frequencies of sunlight from Equation~\ref{numeas}:

\begin{equation}
      \nu_{{\rm meas},m} = \nu_{\rm emit}\left(\frac{(1+z_{{\rm GR},\odot})}{\gamma_\odot (1+\vec{\beta}_\odot\cdot \hat{\rho}_{\odot,m})}\right)\left(\frac{\gamma_m (1+\vec{\beta}_m\cdot \hat{\rho}_{\odot,m})}{(1+z_{{\rm GR},m})}\right)
\end{equation}
and similarly for an observer on Earth of that Solar System object we have:
\begin{equation}
    \nu_{\rm meas} = \nu_{{\rm emit},m}\left(\frac{(1+z_{{\rm GR},m})}{\gamma_m (1+\vec{\beta}_m\cdot \hat{\rho}_{m,\earth})}\right)\left(\frac{\gamma_\earth (1+\vec{\beta}_\earth\cdot \hat{\rho}_{m,\earth})}{(1+z_{{\rm GR},\earth})}\right)
\end{equation}

Since the Doppler shift imparted by the Solar System object in its own frame is zero, we have $\nu_{{\rm meas},m} = \nu_{{\rm emit},m}$. Then the redshifts observed at Earth are
\begin{equation}
    z_{\rm meas} = \frac{\nu^\prime}{\nu_{\rm emit}}\left[\frac{(1+z_{{\rm GR},\earth})\gamma_\odot (1+\vec{\beta}_\odot\cdot \hat{\rho}_{\odot,m})(1+\vec{\beta}_m\cdot \hat{\rho}_{m,\earth})}{(1+z_{{\rm GR},\odot})\gamma_\earth (1+\vec{\beta}_m\cdot \hat{\rho}_{\odot,m})(1+\vec{\beta}_\earth\cdot \hat{\rho}_{m,\earth})}\right]-1
    \label{zmeas_vesta}
\end{equation}

The ``barycentric correction'' in this case is
\begin{equation}
    z_B = \frac{\gamma_\earth (1+\vec{\beta}_\earth\cdot \hat{\rho}_{m,\earth})}{(1+z_{\rm GR,\earth})}
          \frac{(1+\vec{\beta}_m\cdot \hat{\rho}_{\odot,m})}{(1+\vec{\beta}_m\cdot \hat{\rho}_{m,\earth})}
          -1
\end{equation}

Note that this corrects measurements to the vantage of the {\it Solar System object}, not the Earth.  That is, the dominant signal in the barycentric-corrected velocities of Ceres will be the Sun's motion due to Jupiter in the direction of Ceres, modulated at their synodic period.

\subsection[]{Calculation of Retarded Times}

\label{sec:retarded}
For reflected light observations, we need to take into account the light travel time. One can do this iteratively to estimate the position and velocity of the reflected light observation target, until the calculated Doppler shift converges. In our code implementation (Section \ref{code}), we obtain the light travel time directly from our Horizons call for the solar system object, where we first query for the position of the solar system object with respect to the observatory at the time of exposure (\textit{t}) which also gives us the light travel time. We use this light travel time to calculate the retarded position of the target ($t_m$); and then perform a similar step to calculate the light travel time from the solar system object and the Sun, to obtain the time of emission ($t_e$) from the Sun.

\subsection{Importance of light travel time, finite source effects, inelastic scattering, and general relativistic effects}

\label{complexities}
Fortunately, Solar System objects act as good diffuse reflectors and so provide disk-integrated sunlight much the same way telescopes collect disk-integrated starlight. There are some drawbacks to observing them, however.

Unlike with the Sun, using the retarded position of the Solar System object may be important. The Doppler shift error from neglecting this term is of order $v_1v_2/c$, where both velocities in this case are of order 20 km/s, implying typical errors of order $\sim 1$ m/s. These light travel times thus need only be precise to 1 part in 100 or so for 1 cm/s precision, so approximating light travel times using the instantaneous positions of the bodies at the time of observation is sufficient.

Also unlike the Sun, objects such as Ceres, Vesta, and Pallas can have highly variable surface brightness and phases, which cause asymmetric rotational broadening \citep{Lanza15,Molaro16,Haywood16}. Because they are not point sources they fill the telescope pupil differently from stars and so are not perfect proxies for sun-as-a-star \citep{Molaro08}. Also, observing at true opposition can create unexpected effects \citep{Molaro15}.  

Inelastic scattering may also be a concern. Observations of planetary atmospheres can suffer from the the Ring effect \citep{RingEffect}, caused by Raman, Brillouin, and other forms of scattering. To first order, the effect is to ``fill in'' the lines of the solar spectrum \citep{Pallamraju00}, but a wavelength shift is also present \citep{Cochran81}. This effect (to say nothing of wind) complicates attempts to measure the Doppler velocity of the Sun using blue sky measurements \citep{Gray00,Molaro08}. We are not aware of a documented case of a similar effect from planetary surfaces, but if one is present it would be a function of scattering angle and so present additional complexities.

\citet{bary} considered the Shapiro delay, caused by the general relativistic distortion of spacetime as starlight passes by the Sun or Jupiter, because its time derivative (due to the motions of the observer and the source of the delay) introduces a small Doppler shift. For stars within tens of degrees of the Sun, the effect can matter at the mm s$^{-1}$ level, and for relatively bright Solar System objects even closer to the Sun it may matter even more.  The effect is even smaller for Jupiter, even when stars are within 1$^\prime$ of it, but the Galilean satellites can approach even closer than that.  

We have not included this effect in our calculations or in {\tt barycorrpy} because it is small and application of Equation 23 of \cite{bary} to correct for it is nontrivial: the effect depends not only on the angular separation between the Sun (or Jupiter) and the relfecting body, but on their orientation (it should be relevant only when the reflecting object is {\it behind} the Sun or Jupiter), and needs to be considered for both legs of the journey to Earth. Work near or below cm s$^{-1}$ precision may need to include a small correction for sources near the Sun or Jupiter.

\section{Code Implementation}

\label{code}
We have modified the {\tt barycorrpy} package \citep{BARYCORRPY} to handle observations of the Sun directly and in reflection. Its primary function remains the same: given times of observation and measured redshifts ($z_{\rm meas}$), it returns velocities corrected for the motion of the observatory ($z_{\rm true})$. Users may still request that the function instead return $z_B$ the usual way by setting $z_{\rm meas}=0$.

We have added an optional {\tt SolSystemTarget}  parameter that, when set to `Sun' or valid Horizons ID of a Solar System body, applies the analogous correction for sunlight (i.e.\ applying Eq.~\ref{corrected}). In this mode, the code queries  JPL Horizons \citep{JPLHorizons} using the \texttt{astropy} package in \texttt{Python}. The reflected light functionality allows the user to specify the \texttt{HorizonsID\char`_type} parameter for the \texttt{barycorrpy} call, which defaults to `smallbody' for asteroid observations, but can be set to `majorbody' for lunar or planetary observations. Following the Horizons documentation for valid object IDs, the {\tt SolSystemTarget} parameter can be constructed to specify particular coordinates on solar system bodies with good rotational ephemerides. We provide an example of this functionality in the \texttt{barycorrpy} documentation.

Finally, we have added a new functionality to the code that allows users to synthesize radial velocity measurements from a star or the Sun. We call this ``prediction mode'' by analogy with the similarly named mode in {\tt TEMPO} \citep{TEMPO2}.  This produces the redshifts one expects to measure from a target, and can be toggled on by setting the optional keyword \texttt{predictive} to be \texttt{True}. For stars, this is equivalent to the expected value of $z_{\rm meas}$ in the case where $z_{\rm true}=0$ (i.e.\ the case where the star has no orbiting companions), so the code returns 
\begin{equation}
\begin{split}
    z_{\rm predicted} = \frac{1}{1+z_B}-1 & \hspace{40pt}\text{(for stars)}
\end{split}
\end{equation}
In the case of the Sun we can include its precisely known motions due to Solar System objects, so instead the code returns 
\begin{equation}
\begin{split}
    z_{\rm predicted} = z_{\rm meas} & \hspace{60pt}\text{(for the Sun)}
\end{split}
\end{equation}
\noindent where $z_{\rm meas}$ is given by Eq.~\ref{zmeas_sun} or \ref{zmeas_vesta}, as appropriate.  

Our code does not account for phases, uneven albedos, and rotational contributions from unresolved objects. Specifying an object such as ``Ceres'' will return the barycentric correction for mirror at the body center of the specified object. The code will, however, correctly handle resolved observations of bodies such as the Moon because the JPL emphemeris includes orientation information for many Solar System bodies with IAU rotational models. To use this feature, users should specify the surface coordinates of the part of the body they observed using the topocentric coordinate syntax of Horizons.

\section{Precision of the JPL Ephemeris}

\label{sec:precision}

For direct measurements of the Sun, the dominant source of uncertainty is the position of the Sun with respect to Earth in the barycentric frame. The uncertainty of the position of the Sun in the RA direction $\sigma_{\rm RA}$ is order a few hundreds of meters, corresponding to a radial velocity uncertainty of $\sim v_\earth \sigma_{\rm RA}$, which is a few tens of $\upmu$ s$^{-1}$.

In reflected light, the situation is similar. The positional uncertainties of Vesta and Ceres were of order 10 km, but improved by a factor of $\sim 100$ with the arrival of Dawn \citep{Konopliv14,Konopliv18}, and so have similar barycentric uncertainties to direct sunlight.  Observations of other asteroids would have correspondingly larger barycentric correction errors, of order mm s$^{-1}$. 

The positional uncertainties in the RA direction of the rocky planets are highly variable, being of order 1 km for Mercury, 100 m for Venus, 10 km for Mars and Jupiter, and around few km for Saturn. In all cases, the current barycentric correction error is of order mm s$^{-1}$ or better \citep{Folkner11}.  These uncertainties grow with time as the ephemerides become stale, but shrink dramatically when spacecraft such as Juno arrive that can provide precise ranging data.

The JPL Ephemerides are computed in a manner that optimizes the uncertainties in the positions and velocities of certain solar system objects at times of interest for planetary science work by NASA. These ephemerides also do not provide rigorous uncertainties. Because the JPL ephemerides are sufficiently precise for current EPRV work in the Solar System, we use them here, but more rigorous treatment of uncertainties is possible.  For instance, \citet{Vallisneri20} computed a custom ephemeris product for the detection of gravitational waves via pulsar timing, and \citet{Cionco18} produced a custom ephemeris to precisely study the barycentric motion of the Sun.

\section{Detecting Solar System Planets via Precise Doppler Measurements of the Sun}

\label{sec:detecting}

\subsection{From Direct Observations of the Sun}

Using our code, we have simulated observations of the Sun every 10 days from Cerro Tololo Inter-American Observatory for 30 years to explore how well the effects of the Solar System planets can be inferred from Solar RVs (as $z_{\rm predicted}$, above). We chose this uniform cadence because it is much faster than any relevant frequencies in the problem, and so will not generate any aliased power in our periodograms.

\begin{figure}
    \plotone{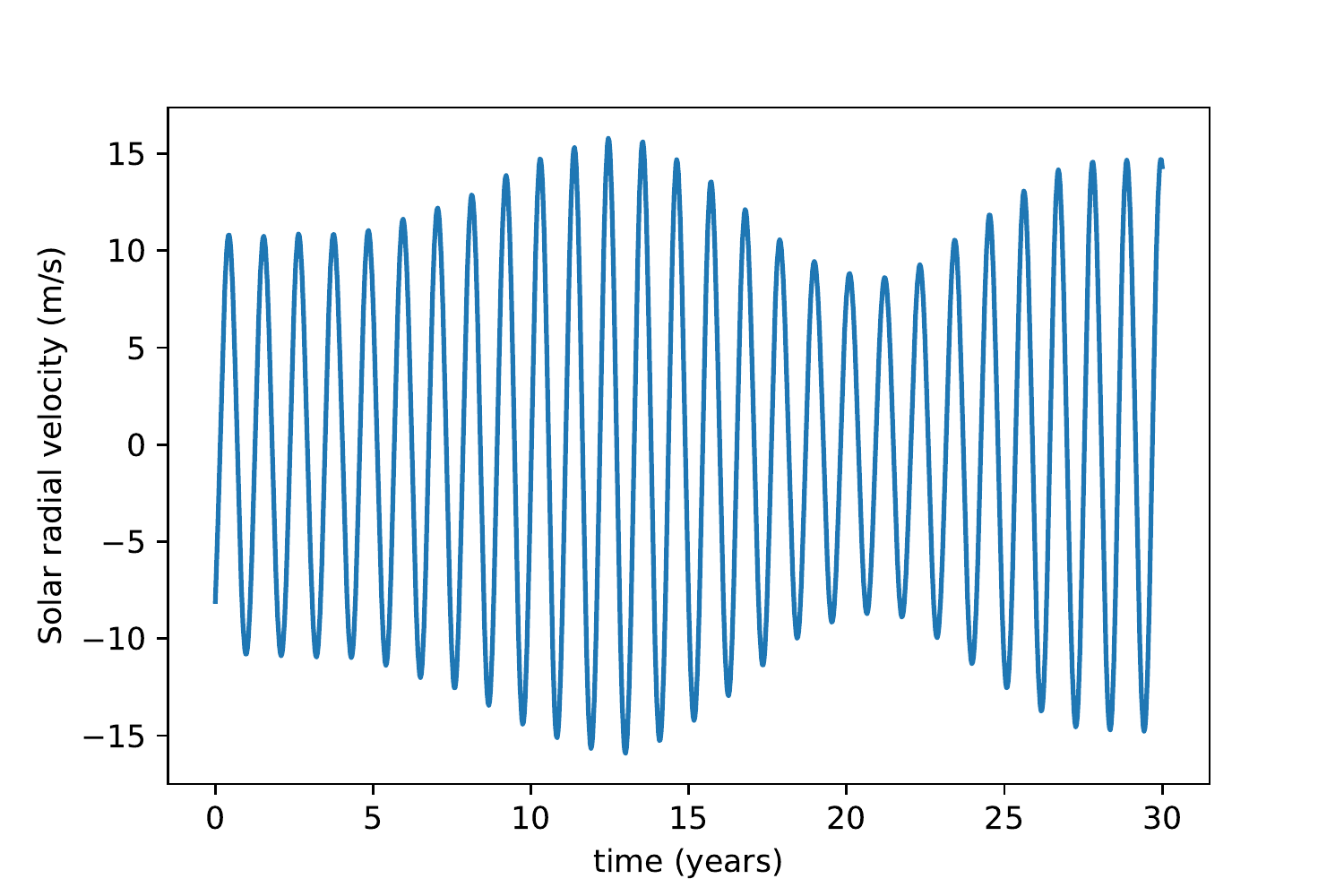}
    \caption{Motion of the Sun in the direction of the Earth after barycentric correction. The dominant signal is that of the motion of the Sun due to Jupiter, modulated by the orbital motion of the Earth, at the synodic period of Jupiter ($\sim$ 400 days). The longer period modulation is presumably due to the orbits of Jupiter and Saturn.}
    \label{fig:RV}
\end{figure}

Applying the barycentric correction formalism above to determine $z_{\rm true}$ for these observations, we can isolate only the motion of the Sun in the direction of the observatory, having removed all Doppler components of the observatory motion from the signal.  Figure~\ref{fig:RV} shows this result, which has a strong signal at the synodic period of Jupiter with amplitude of $\sim 10$ m s$^{-1}$. 

This is not surprising, since Jupiter is responsible for most of the Sun's motion, and we observe an ever-changing perspective on that motion at that synodic period.  

Since our synthetic data are noise-free and periodically sampled, we can take a Fourier transform to determine the ideal periodogram one could achieve in 30 years daily observations.  

\begin{figure}
    \plotone{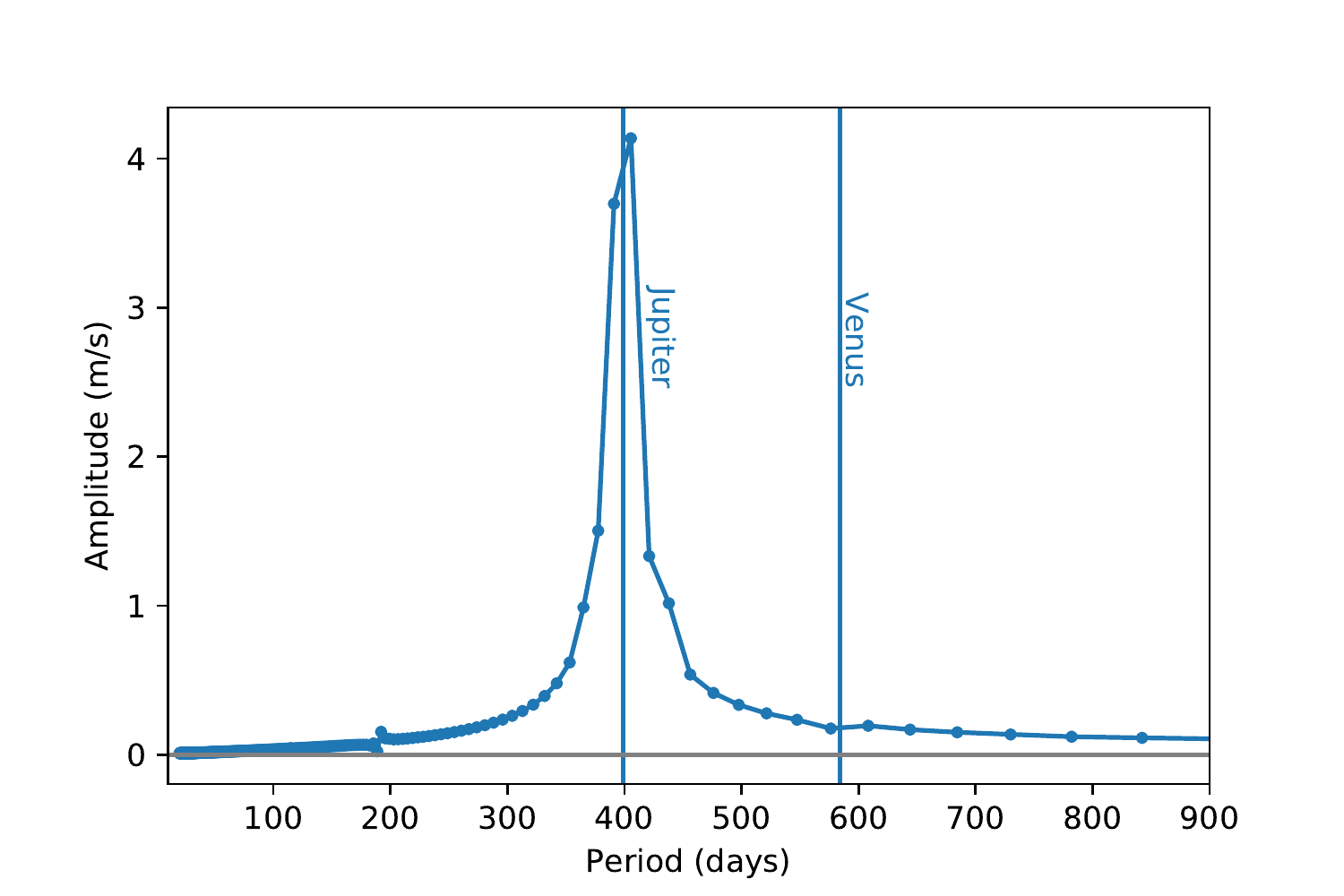}
    \caption{Fourier transform of Figure~\ref{fig:RV}, expressed as semiamplitude as a function of period. The synodic periods of Jupiter and Venus are indicated. The power from Jupiter's influence is broadly distributed, and completely overwhelms the signal of Venus. Improving the frequency resolution would require extending the baseline of observations beyond 30 years.  The feature at 190 days at the synodic period of Jupiter and half of Earth's period (i.e.\ it appears at frequency $2\nu_\earth+\nu_{\rm Jupiter}$), and is caused by the eccentricity of Earth's orbit.}
    \label{fig:FT}
\end{figure}

We present the result in Figure~\ref{fig:FT}. The 12 m s$^{-1}$ ``wobble'' of the Sun due to Jupiter is seen to be broadly distributed across a range of frequencies with a peak at the synodic period of Jupiter.  The width of this signal is likely due to the combination of the significant eccentricity of Jupiter (e=0.049) and the non-commensal periods of Jupiter and Earth, which conspire to generate a non-periodic RV signal. That is, in the barycentric frame co-rotating with the Earth, Jupiter's orbit is neither circular nor periodic, and so does not produce power at a single frequency and its harmonics.

The wide wings of this power extends well into the synodic periods of the other planets.  There is no peak at the synodic period of Venus, the next most significant planet in the time series, presumably because it is buried beneath the broad signal of Jupiter.

It is therefore quite challenging to ``detect'' the presence of any planet other than Jupiter (and possibly Saturn) even with perfect, daily measurements of the Sun for 30 years. One would need to first subtract the signal of Jupiter to recover that of Venus, the next strongest signal, but this is not straightforward and requires carefully choosing a counterfactual of the sort discussed in Section~\ref{sec:challenges}.  

For instance, one might construct an artificial ephemeris by using the true positions of Jupiter and the Sun from the JPL ephemeris, but ascribing to the Sun velocities calculated from accelerations due only to Jupiter. Because the indirect effects of Venus on the positions of Jupiter and the Sun are small, this is likely a sufficiently precise ephemeris to feed {\tt barycorrpy} in predictive mode such that the only remaining signal is the radial motion of the Sun due to the planets other than Jupiter.  The signal of Venus could then be similarly calculated and searched for in the residuals via a matched filter, or in a manner analogous to exoplanet hunting via model fitting, perhaps with Bayesian and MCMC methods.

To use Solar RV measurements to prove the precision of an RV spectrograph to terrestrial, Habitable Zone planets it would be simpler, perhaps, to subtract the known signal of all of the planets and perform an injection-recovery test to see if one would be sensitive to Venus if one observed the Sun from a fixed vantage. 

\subsection{From Observations of the Sun in Reflection}

\begin{figure}
    \plotone{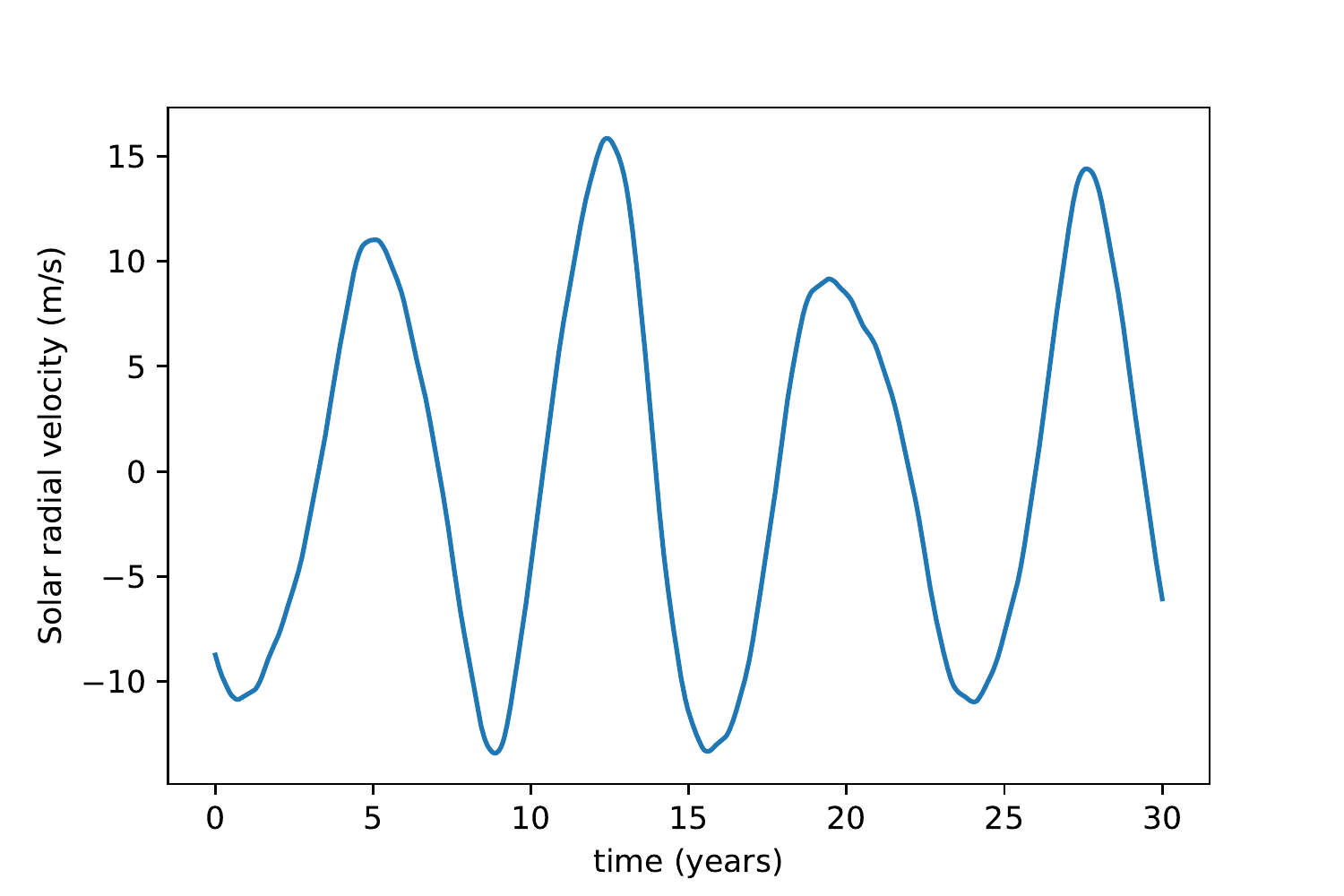}
    \caption{Radial velocity of sunlight reflected from Ceres as observed every 10 days from Earth over 30 years, after barycentric correction. The signal from Jupiter stands out clearly, but the eccentricities of it and Ceres make for a complex waveform.}
    \label{fig:Ceres_RV}
\end{figure}

We have also computed the radial velocities one would measure in reflection from Ceres at the same times. Since our purposes here are purely illustrative, we have calculated these velocities every 10 days for 30 years without regard to the true observability of Ceres, or to the complexities described in Section~\ref{complexities} \citep[e.g.][]{Molaro15}. The raw radial velocities are dominated by the diurnal motion of the Earth, but the barycentric-corrected velocities show structure at many long frequencies (Figure~\ref{fig:Ceres_RV}). The signal of Jupiter is clearly visible, but there is also significant power at other frequencies, likely due to the significant eccentricity in the orbit of Ceres.

\begin{figure}
    \plottwo{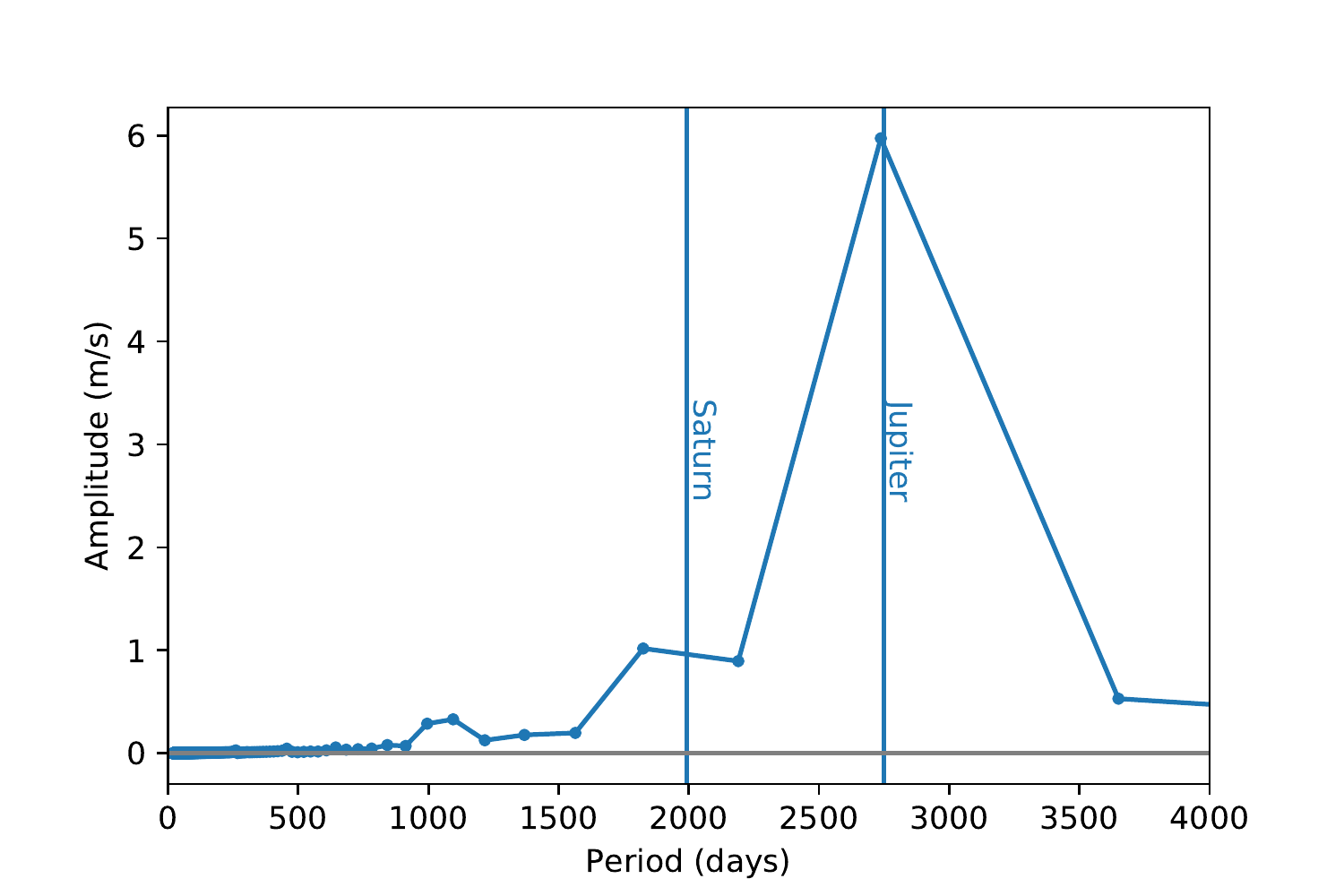}{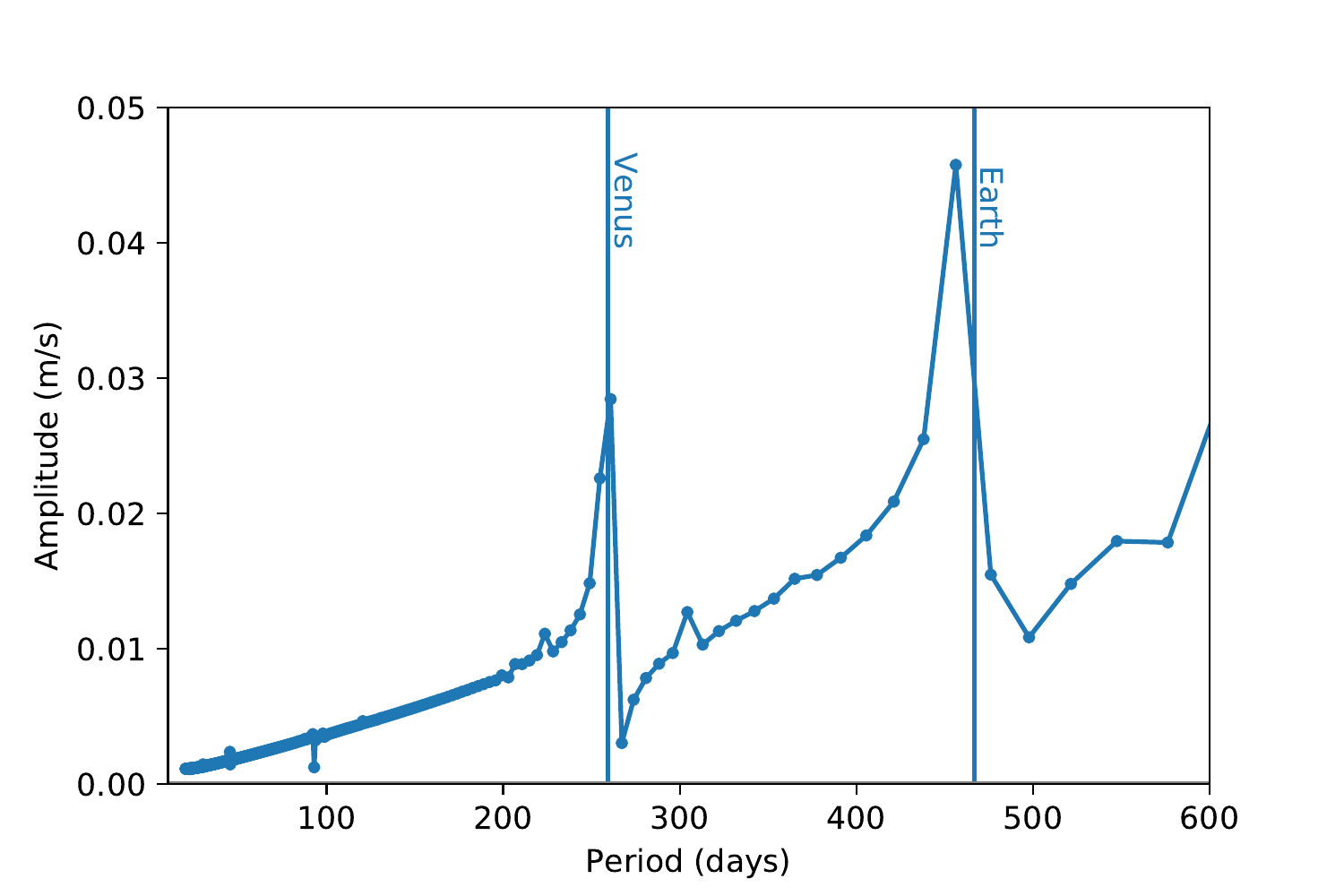}
    \caption{Fourier transform of Figure~\ref{fig:Ceres_RV}, expressed as semiamplitude as a function of period. {\it Left}: The synodic periods of Jupiter and Saturn are indicated. Improving the frequency resolution would require extending the baseline of observations beyond 30 years. {\it Right}: Detail, with synodic periods of Earth and Venus are indicated. These periods are sufficiently distinct from those of Jupiter and sufficiently high frequency that they stand out cleanly.
    }
    \label{fig:Ceres_FT}
\end{figure}

Because the synodic period between Jupiter and Ceres is so long (Ceres has an orbital period of 4.6 years) even after 30 years there is very little frequency resolution in the resulting Fourier transform, shown in Figure~\ref{fig:Ceres_FT}, making it difficult to determine whether the signal of Saturn is present.

Interestingly, the synodic periods of Earth and Venus in these data sets are sufficiently distinct from that of Jupiter that their signals clearly stand out in these noiseless data, as seen in Figure~\ref{fig:Ceres_FT}. They are not sharply peaked, however, presumably due to the orbital eccentricity of Ceres. This suggests that if the difficulties in making precise radial velocity measurements of Ceres (or other asteroids) can be overcome, that a sufficiently dense and long time series of them might be a good proxy for detecting Habitable Zone terrestrial planets.

\section{Summary and Conclusions}

We have derived and explained a barycentric correction procedure for Doppler measurements of sunlight, and provided it as part of the public Python code {\tt barycorr}.  We have included both the case of direct observations of the Sun and for sunlight in reflection off of solar system objects.

We have discussed the philosophical challenges to defining a ``barycentric correction'' in the context of sunlight, and offered two solutions: to provide a model for the expected Doppler shift of sunlight for a given observatory at a given time, and to reduce measurements to only the component of the Sun's motion in the direction of the observatory at the time of observation.

We find that the JPL ephemerides, as provided by the Horizons service, are sufficiently precise for EPRV work that uses the Sun as a ``standard star'' for determining instrumental precision and studying astrophysical sources of RV noise. 

We have used our code to explore the challenges of ``detecting'' the Solar System planets in data from direct observations of the Sun, and find that while the signal of Jupiter stands out cleanly, the signals of Earth and Venus are not distinct in Fourier space even in evenly sampled, noiseless data. We do find, however, that Earth and Venus have small but distinct signals in Fourier space when the Sun is observed in reflection, for instance from Ceres.

\begin{acknowledgements}

{We thank William Folkner for helping us validate our formulae and our interpretation of the JPL ephemeris by using custom software to compute precise arrival frequencies of hypothetical Solar System spacecraft for comparison with {\tt barycorrpy}. We also thank Andrea Lin, Paolo Molaro, Christophe Lovis, Xavier Dumusque, Fabo Feng, and Antonino Lanza for helpful discussions and feedback on this paper, and for their previous work on the subject. We thank the anonymous referees for their helpful feedback.

Section~\ref{sec:intro} is taken largely from a telescope proposal written by JTW, and incorporates suggestions from our co-investigators Andrew Collier Cameron, Heather Cegla, Rapha\"elle Haywood, and Jayadev Rajagopal.

This research has made use of NASA's Astrophysics Data System Bibliographic Services and Astropy,\footnote{\url{http://www.astropy.org}} a community-developed core Python package for Astronomy \citep{astropy}.

The Center for Exoplanets and Habitable Worlds and the Penn State Extraterrestrial Intelligence Center are supported by the Pennsylvania State University and the Eberly College of Science.}

\end{acknowledgements}

\bibliography{references}

\end{document}